\documentclass[useAMS,usenatbib]{mn2e}
\usepackage{epsfig,rotate,graphicx}
\usepackage[fleqn]{amsmath}
\usepackage{subfigure}
\usepackage{lscape}
\usepackage{bm}

\usepackage{amsmath}
\usepackage{bm}
\newcommand{\p}{\partial}
\newcommand{\lmax}{l_\mathrm{max}}
\newcommand{\sbar}{\bar{\sigma}} 
\newcommand{\imgi}{\mathrm{i}}

\newcommand{\avg}[1]{\langle{#1}\rangle}

\newcommand{\real}{\operatorname{Re}}

\newcommand{\zmax}{Z_s}
\defcitealias{lin12}{L12} 

\newcommand{\mnras}{MNRAS}
\newcommand{\apj}{ApJ}
\newcommand{\aap}{A\&A}
\newcommand{\apjl}{ApJL}

\newcommand{\dd}{\delta}

\newcommand{\be}{\begin{equation}}
\newcommand{\ee}{\end{equation}}
\newcommand{\gtrsim}{\;\raisebox{-.8ex}{$\buildrel{\textstyle>}\over\sim$}\;}

\newcommand{\pasj}{{\it PASJ, }}

\title[Linear RWI in 3D]{Effects of upper disc boundary conditions on the
  linear Rossby wave instability}

\author[Lin]{ Min-Kai Lin
  \thanks{E-mail: mklin924@cita.utoronto.ca} \\
Canadian Institute for Theoretical Astrophysics,
60 St. George Street, Toronto, ON, M5S 3H8, Canada \\
}

\begin{document}

\maketitle
\begin{abstract}

The linear Rossby wave instability (RWI) in global, three-dimensional (3D)
polytropic discs is revisited with a much simpler numerical method
than that previously employed by the author. The governing partial
differential equation is solved with finite-differences in the radial
direction and spectral collocation in the vertical
direction. RWI modes are calculated subject to different upper
disc boundary conditions. These include free surface, solid boundaries and
variable vertical domain size. Boundary conditions that oppose vertical
motion increases the instability growth rate by a few per cent. 
The magnitude of vertical flow throughout the fluid column can be affected
but the overall flow pattern is qualitatively unchanged. Numerical
results support the notion that the RWI is intrinsically 
two-dimensional. This implies that inconsistent upper disc
boundary conditions, such as vanishing enthalpy perturbation, may
inhibit the RWI in 3D.     

\end{abstract}

\section{Introduction}
The Rossby wave instability \citep[RWI,][]{lovelace99,li00} is the 
thin-disc analog of the Papaloizou-Pringle instability in thick tori
\citep{papaloizou84,papaloizou85,papaloizou87}. The RWI is a shear 
instability associated with extrema in the potential vorticity profile of a disc. 
Its occurrence in radially structured protoplanetary discs, leading  to vortex 
formation \citep{li01}, can transport angular momentum, 
assist planet formation \citep{lyra08, lyra09} and affect planetary
migration \citep{li09,lin10}.       

The RWI was originally described in two-dimensional (2D) discs. Recent works
show that it can operate in 
three-dimensional (3D) geometry \citep{umurhan10, meheut10,
  meheut12b,meheut12,lin12}. In these calculations the perturbation 
extends vertically through the disc, but the effect of    
conditions at the disc surface was not investigated. This might be a
relevant issue for the RWI to develop in non-magnetized regions of
protoplanetary discs \citep{lyra12}. In the layered accretion scenario,
this `dead zone' is overlaid by an actively accreting layer
\citep{gammie96,okusumi11,martin12}. A first step towards revealing  
vertical boundary effects on the 3D RWI is to calculate it 
subject to different upper disc boundary conditions. 

Previous linear RWI simulations assume the zero-density surface as the
upper disc boundary. When the density stratification has the
appropriate functional form, regularity conditions enable the
vertical dependence of perturbations to be expressed in terms of classic
orthogonal polynomials \citep{papaloizou85, takeuchi98}. The linear
problem becomes a set of ordinary differential equations 
(ODE), but these ODEs can be complicated. 
Furthermore, this approach does not permit boundary conditions at other heights to be 
imposed \citep[][hereafter \citetalias{lin12}]{lin12}.

The purpose of this work is to address the above caveat of
\citetalias{lin12}. An alternative numerical method is applied to the
linear problem so that effects of upper disc boundary conditions can be
explored through simple numerical experiments. This paper is organized as
follows. \S\ref{setup} defines the equilibrium and perturbed disc. 
\S\ref{numerical_method} gives a 
step-by-step description to convert the problem into a single matrix
equation. Solutions are presented in
\S\ref{results} and discussed in \S\ref{summary}.

\section{Model and governing equations}\label{setup}
The system is a geometrically thin, non-self-gravitating, inviscid 
fluid disc orbiting a central star of
mass $M_*$. The disc is governed by the Euler equations in $(r,\phi,z)$
cylindrical co-ordinates centered on the star. Units such
that the gravitational constant $G=M_*=1$ are adopted. A
polytropic equation of state is assumed, so that the pressure $P$ is
related to the density $\rho$ by $  P = K \rho^{1+1/n},$ where $n$ is
the polytropic index and $K$ is a constant.   

Details of the equilibrium disc are given in 
\citetalias{lin12}. The unperturbed disc is steady and axisymmetric with 
surface density profile in the form $\Sigma\propto r^{-\alpha}B(r)$,
where $B(r)$ describes a Gaussian bump centered at $r=r_0$ with
amplitude $A$ and width $\Delta r$. The density is 
$\rho(r,z)=\rho_0(r)[1-z^2/H^2(r)]^n$, where $\rho_0$ is the midplane density
and $H(r)$ is the disc thickness. The unperturbed velocity field is 
$(v_r,v_\phi,v_z)=(0,r\Omega,0)$ where $\Omega(r)$ is the angular
velocity, and $\Omega\simeq\Omega_k\equiv\sqrt{GM_*/r^3}$. 
Fiducial parameter values are: $\alpha=0.5,\,r_0=1,
A=1.4,\,\Delta r=0.05r_0$ and $H(r_0)=0.14r_0$. 

Eulerian perturbations to the above equilibrium in the form
$\real[\dd\rho(r,z)\exp{\imgi(\sigma t + m\phi)}]$ are considered.   
$\sigma=-(\omega+\imgi\gamma)$ is a complex eigenfrequency and $m$ is 
a positive integer. $\gamma$ and $-\omega$ are the  mode growth rate
and real frequency, respectively. The linearized fluid equations yield
the following partial differential equation (PDE):    
\begin{align}\label{3d_linear_poly}
  r\dd\rho =&  \frac{\p}{\p r}\left(\frac{r\rho }{D} \frac{\p W}{\p
    r}\right) + \frac{2mW}{\sbar} 
  \frac{\p}{\p r}\left(\frac{\rho\Omega}{D}\right) \notag\\
  &- \left(\frac{m^2\rho}{rD}\right) W
  -\frac{r }{\sbar^2}\frac{\p}{\p z}\left(\rho\frac{\p W}{\p
    z}\right) 
\end{align}
\citepalias{lin12}, where $W\equiv\dd P/\rho$ is the enthalpy
perturbation and $D\equiv \kappa^2 - \sbar^2$; where $\kappa^2\equiv
r^{-3}d(r^4\Omega^2)/dr$ is the square of the epicycle frequency and
$\sbar = \sigma + m\Omega$ is the shifted frequency. The RWI is
associated with an extremum in potential vorticity
\citep[$\eta\equiv\kappa^2/2\Omega\Sigma$,][]{lovelace99}. 
In the above  setup $\mathrm{min}(\eta)$ 
occurs near $r_0$, so that $\sbar(r_0)\sim0$.
The aim is to solve Eq. \ref{3d_linear_poly} as a PDE eigenvalue problem
with various boundary conditions applied at $z/H = \zmax$. 

For
simplicity $\zmax$ is assumed constant, but
extension to $\zmax=\zmax(r)$ is straight forward provided that
$\zmax(r)\neq0$   
\citep[in which case transformation to polar-like coordinates is needed, see][]{papaloizou84}. 

\subsection{Vertical boundary conditions}\label{vbc}
The enthalpy perturbation is assumed to be symmetric about the
midplane, since even modes have been found to dominate in nonlinear
simulations \citep{meheut10}. The upper disc boundary is set to $\zmax
< 1$, thus avoiding the disc surface where the PDE becomes
singular. The following upper boundary conditions (BC) were found to
permit the RWI.   

\emph{`Open' boundary}. The Lagrangian pressure
perturbation vanishes at $\zmax$:
\begin{align}
  \Delta P 
  &\equiv \dd P + \bm{\xi}\cdot\nabla P=0
\end{align}
where $\bm{\xi}$ is the Lagrangian displacement. 
If $\zmax$ is unity
then this is the usual condition for polytropes with vanishing pressure at 
its surface (which is satisfied automatically for regular solutions).   

\emph{`Solid' boundaries.} At $\zmax$ the meridional velocity
perturbation satisfies
\begin{align}
  \dd v_z = \nu\dd v_r, 
\end{align}
where $\nu=\nu(r)$ is a real function. Choosing $\nu = 
\zmax dH/dr$ corresponds to zero velocity perpendicular to the upper
disc boundary ($\dd v_\perp=0$). This applies to an impermeable
surface. The simpler choice $\nu\equiv0$ corresponds to zero vertical
velocity ($\dd v_z=0$). This might apply, for example, when vertical
velocities are strongly damped above $\zmax$. In practice, the two 
conditions are similar because $|\nu|\ll1$ for a thin disc.  

\section{Numerical method}\label{numerical_method}
It is convenient to adopt the co-ordinate system $(R,Z) \equiv
(r,z/H)$ so the computational domain is rectangular for constant
$\zmax$. Then the governing equation becomes  
\begin{align}\label{gov_eq}
& 0 =  \mathcal{A}\frac{\p^2W}{\p R^2} + \mathcal{B} \frac{\p^2W}{\p
    Z\p R} + \mathcal{C}\frac{\p^2W}{\p Z^2} 
  +\mathcal{D}\frac{\p W}{\p R} + \mathcal{E}\frac{\p W}{\p Z} +
  \mathcal{F}W,\\
&  \mathcal{A} = (1-Z^2)R^2, \\
&  \mathcal{B} = -2Z(1-Z^2)R^2\frac{H^\prime}{H}, \\
&  \mathcal{C} = Z^2(1-Z^2)R^2\left(\frac{H^\prime}{H} \right)^2 -
  \frac{R^2D(1-Z^2)}{\sbar^2H^2}, \\
&  \mathcal{D}  = \left[\ln{\left(\frac{\rho_0
        R}{D}\right)}\right]^\prime(1-Z^2)R^2 +
  2nZ^2R^2\frac{H^\prime}{H}, \\
&  \mathcal{E} = 2nZ\frac{R^2D}{\sbar^2H^2}
  +ZR^2\left(\frac{H^\prime}{H}\right)^2\left[1-(2n+1)Z^2\right] \notag\\
&- Z(1-Z^2)R^2\left\{
  \left(\frac{H^\prime}{H}\right)^\prime + \left(\frac{H^\prime}{H}\right)\left[\ln{\left(\frac{\rho_0
        R}{D}\right)}\right]^\prime\right\}, \\
&  \mathcal{F}  = \frac{2mR\Omega}{\sbar}\left[\ln{\left(\frac{\rho_0
        \Omega}{D}\right)}\right]^\prime(1-Z^2) +
  2nZ^2\left(\frac{2mR\Omega H^\prime}{\sbar H}\right) \notag\\
  &- m^2(1-Z^2) -
  \frac{nD\rho_0^{-1/n}R^2}{K(1+n)},
\end{align}
and primes denote differentiation with respect to the argument. 

To construct a numerical scheme, the radial co-ordinate is discretized
into $N_R$ grid points uniformly spaced by $\Delta R$. Let $R_i$ 
denote the radial co-ordinate of the $i^\mathrm{th}$ grid point, define 
$W_i(Z)$ as the solution along the vertical line $R=R_i$ and approximate 
it with a linear combination of basis functions,
\begin{align}\label{cheby_expand}
  W_i(Z) &\equiv W(R_i, Z) 
  = \sum_{k=1}^{N_Z} a_{ki}\psi_k(Z/\zmax),
\end{align}
for $1\leq i\leq N_R$, with  
$  \psi_k  = T_{2(k-1)},  $
where $T_l$ is a Chebyshev polynomial of the first
kind of order $l$ \citep{stegun65}. The number of basis functions is
$N_Z$ and $l_\mathrm{max}=2(N_Z-1)$ is the highest polynomial order.    

The pseudo-spectral coefficients $a_{ki}$  are obtained by demanding
the governing PDE, Eq. \ref{gov_eq}, to be satisfied at vertical grid
points $Z_j$ along each line, where   
\begin{align}
  Z_j = -\zmax\cos{\left[\frac{(j-1+\lmax/2)\pi}{\lmax}\right]}\quad 1\leq j \leq N_Z 
\end{align}
This is a standard choice for collocation points in 
pseudo-spectral methods with Chebyshev polynomials \citep{boyd01}. 
Note that only the upper half plane ($z\geq0$) is covered, by the 
assumption of symmetry. 

The next step is to insert Eq. \ref{cheby_expand} into 
Eq. \ref{gov_eq} and evaluate the governing PDE at each $(R_i,
Z_j)$. Radial derivatives are approximated with 
central differences while vertical derivatives are computed 
exactly. At each grid point the result 
is \begin{align}\label{discretized_gov}  
  0 &=\sum_{k=1}^{N_Z} a_{ki}\times \left[\left( \mathcal{F}_{ij}
    -\frac{2\mathcal{A}_{ij}}{\Delta R^2}\right)\psi_{kj} +
    \mathcal{E}_{ij}\frac{\psi^\prime_{kj}}{\zmax} 
    +\mathcal{C}_{ij}\frac{\psi^{\prime\prime}_{kj}}{\zmax^2}\right]\notag\\
  &+\sum_{k=1}^{N_Z} a_{k,i-1}\times
  \left[\left(\frac{\mathcal{A}_{ij}}{\Delta
    R^2} - \frac{\mathcal{D}_{ij}}{2\Delta R}\right) \psi_{kj} - \frac{\mathcal{B}_{ij}}{2\Delta
    R}\frac{\psi^\prime_{kj}}{\zmax} \right]\notag\\
  &+\sum_{k=1}^{N_Z} a_{k,i+1}\times
  \left[\left(\frac{\mathcal{A}_{ij}}{\Delta
    R^2} + \frac{\mathcal{D}_{ij}}{2\Delta R}\right) \psi_{kj} + \frac{\mathcal{B}_{ij}}{2\Delta
    R}\frac{\psi^\prime_{kj}}{\zmax} \right],
\end{align}
where $\mathcal{A}_{ij} = \mathcal{A}(R_i,Z_j)$ and similarly for
other PDE coefficients, and $\psi_{kj}\equiv\psi_k(Z_j/\zmax)$. 
Eq. \ref{discretized_gov} can be written in matrix notation: 
\begin{align}
  \bm{V}^{(i)}\bm{a}^{(i)} + \bm{V}^{(i)}_{-}\bm{a}^{(i-1)} +
  \bm{V}^{(i)}_+\bm{a}^{(i+1)} = \bm{0}.
\end{align} 
The operators $\bm{V}^{(i)}$ and $\bm{V}_\pm^{(i)}$ are $N_Z\times
N_Z$ matrices and the column vector $\bm{a}^{(i)}$ consists of the
$N_Z$ pseudo-spectral coefficients associated with the $i^\mathrm{th}$
line. $\bm{V}_\pm^{(i)}$ represent coupling to adjacent lines. The
matrix elements can be read off Eq. \ref{discretized_gov}. 

For simplicity, $\p_R W = 0$ is set at radial boundaries\footnote{The RWI
typically has disturbance radially confined near $r_0$ and  
therefore insensitive to distant radial boundaries.}. This is achieved 
implicitly by setting  $a_{k,i-1}=a_{k,i+1}$ for 
$i=1,\,N_R$. $\bm{V}^{(1)}_+$ and $\bm{V}_{-}^{(N_R)}$ are 
modified accordingly. At $\zmax$, the governing equation is
replaced by upper disc boundary conditions listed in \S\ref{vbc}. This 
corresponds to modifying the PDE coefficients and the matrix
elements for $j=N_Z$.  

There are $N_R$ lines in total, so the entire system of equations is 
\begin{align}\label{matrix_eq}
   \bm{M}\bm{a} = \bm{0},
\end{align}
where $\bm{M}$ is a  $(N_RN_Z)\times(N_RN_Z)$ block tridiagonal matrix
and $\bm{a}$ is a column vector of length $N_RN_Z$, representing the 
linear operators and the $a_{ki}$, respectively. 
 
The discretized problem, Eq. \ref{matrix_eq}, is a set of linear
homogeneous equations. A solution method is described in 
\citetalias{lin12}. The eigenfrequency 
$\sigma$ is varied, using Newton-Raphson iteration, until 
the matrix $\bm{M}$ becomes singular. Only solutions 
where the reciprocal condition number of $\bm{M}$ is zero 
at machine precision are accepted. \texttt{LAPACK} was used for the
matrix operations, which also yield $\bm{a}$ up to an arbitrary
normalization.  

\section{Linear simulations}\label{results}
The computational domain is $R\in[0.4,1.6]r_0$ and $\zmax=0.9$ or
$\zmax=0.45$. The default polytropic index is $n=1.5$ but cases with
$n=0.7,\,2.5$ are also presented. Linear modes with azimuthal 
wavenumber $m=3$ are considered. 

The fiducial case below employed a trial 
frequency $\sigma = -(0.994m + 0.107\imgi)\Omega_0$ where 
$\Omega_0\equiv\Omega(r_0)$. This was the solution found 
in \citetalias{lin12}. Eigenfrequencies for other 
parameter values may be found by varying 
those parameters slowly away from fiducial values. 
The numerical grid is $(N_R,N_Z)=(384,8)$,   
which corresponds to $\lmax=14$ (cf. $\lmax=6$ 
in \citetalias{lin12}).    

Table \ref{eigen} summarizes the calculations. The last column, 
$\avg{\theta_m}$, is a measure of three-dimensionality of the flow near
$r_0$. $\theta_m$ is defined by 
\begin{align}
  \theta_m^2(r,z) \equiv  \tilde{v}_z^2/(\tilde{v}_z^2 +\tilde{v}_r^2), 
\end{align}
where $\tilde{v}_{r,z}\equiv\real[\dd v_{r,z}W^*(r_0,0)]/|W(r_0,0)|$ is 
the real velocity perturbation at an azimuth close to a local maximum  
in midplane enthalpy perturbation (hereafter the \emph{vortex core}). 
$\avg{\cdot}$ denotes spatial averaging over $R\in[0.8,1.2]r_0$ and
$Z\in[0,\zmax]$.  

\begin{table}
  \caption{Summary of linear simulations.The azimuthal wavenumber is $m=3$ for all cases. } 
    \begin{tabular}{lclcccc}
      \hline
      Case&$n$& $\zmax$ & BC & $\omega/m\Omega_0$ &
      $\gamma/\Omega_0$ & $\avg{\theta_m}$ \\
      \hline\hline
      
      1a &1.5 &0.90 & $\Delta P = 0$ & 0.9941 & 0.1074 & 0.31 \\
      
      1b &1.5 &0.45 & $\Delta P = 0$ & 0.9997 & 0.0889 & 0.36 \\
      
      2a &1.5 &0.90 & $\dd v_\perp = 0$ & 0.9940 & 0.1091&  0.26\\ 
      
      2b &1.5 &0.45 & $\dd v_\perp = 0$ & 0.9939 & 0.1174&  0.06\\
      
      3a &1.5 &0.90 & $\dd v_z =0$ & 0.9945  & 0.1086 & 0.26 \\
      
      3b &1.5 &0.45 &$\dd v_z =0$ &0.9973 & 0.1143 & 0.06 \\
      
      4a &0.7 &0.90 &$\Delta P =0$ &0.9959 & 0.1976 & 0.34 \\  
      
      4b &0.7 &0.90 &$\dd v_z =0$ &0.9968 & 0.2034 & 0.22 \\ 
      
      5a &2.5 &0.90 &$\Delta P =0$ &0.9926 & 0.0359 & 0.39 \\
      5b &2.5 &0.90 &$\dd v_z =0$ &0.9927 & 0.0361 & 0.38 \\
      \hline
    \end{tabular} \label{eigen}
\end{table}

\subsection{Fiducial case}
\begin{figure}
   \centering
   \includegraphics[scale=.425,clip=true,trim=0cm 1cm 0cm
     0cm]{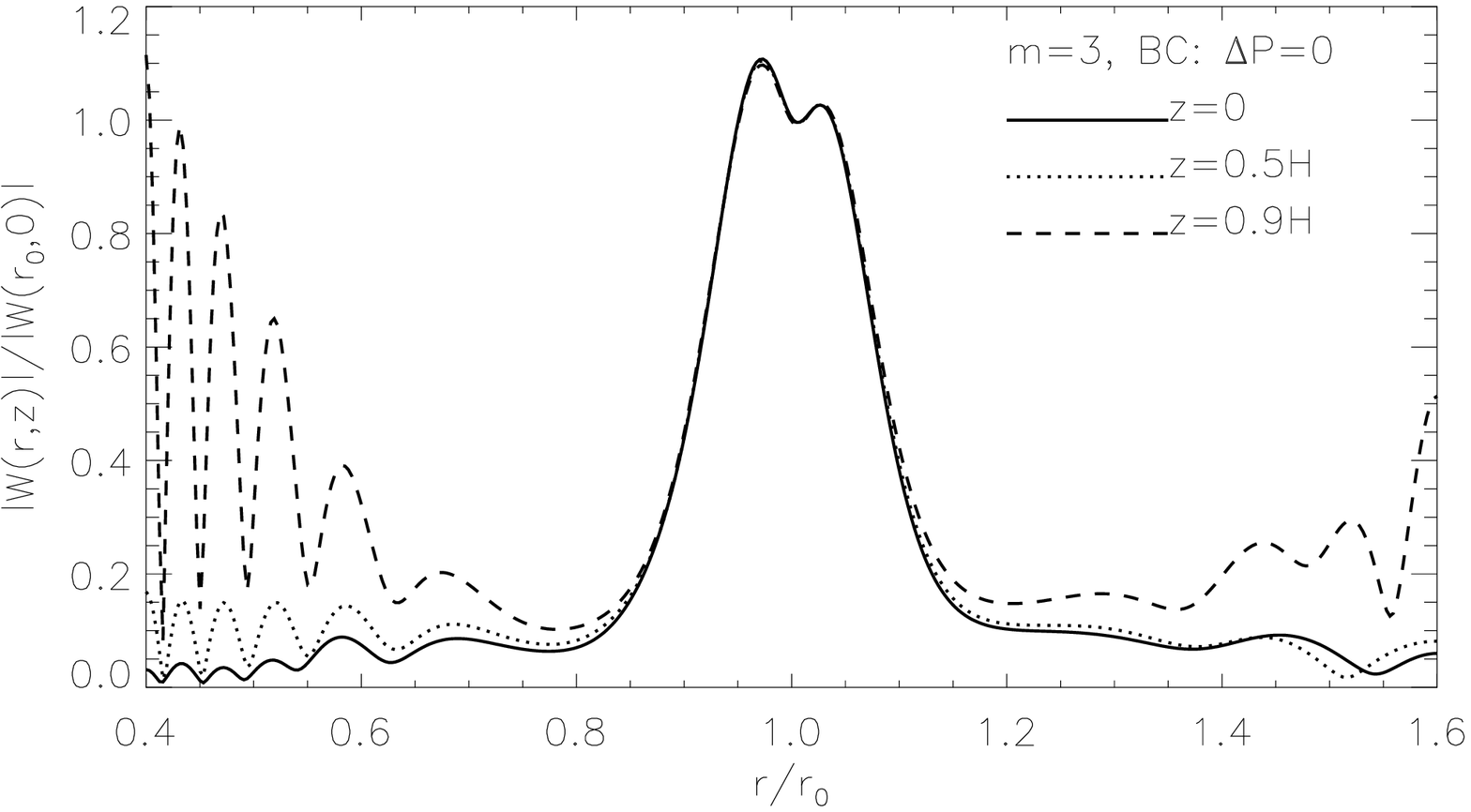} 
    \includegraphics[scale=.425,clip=true,trim=0cm 0cm 0cm
      0cm]{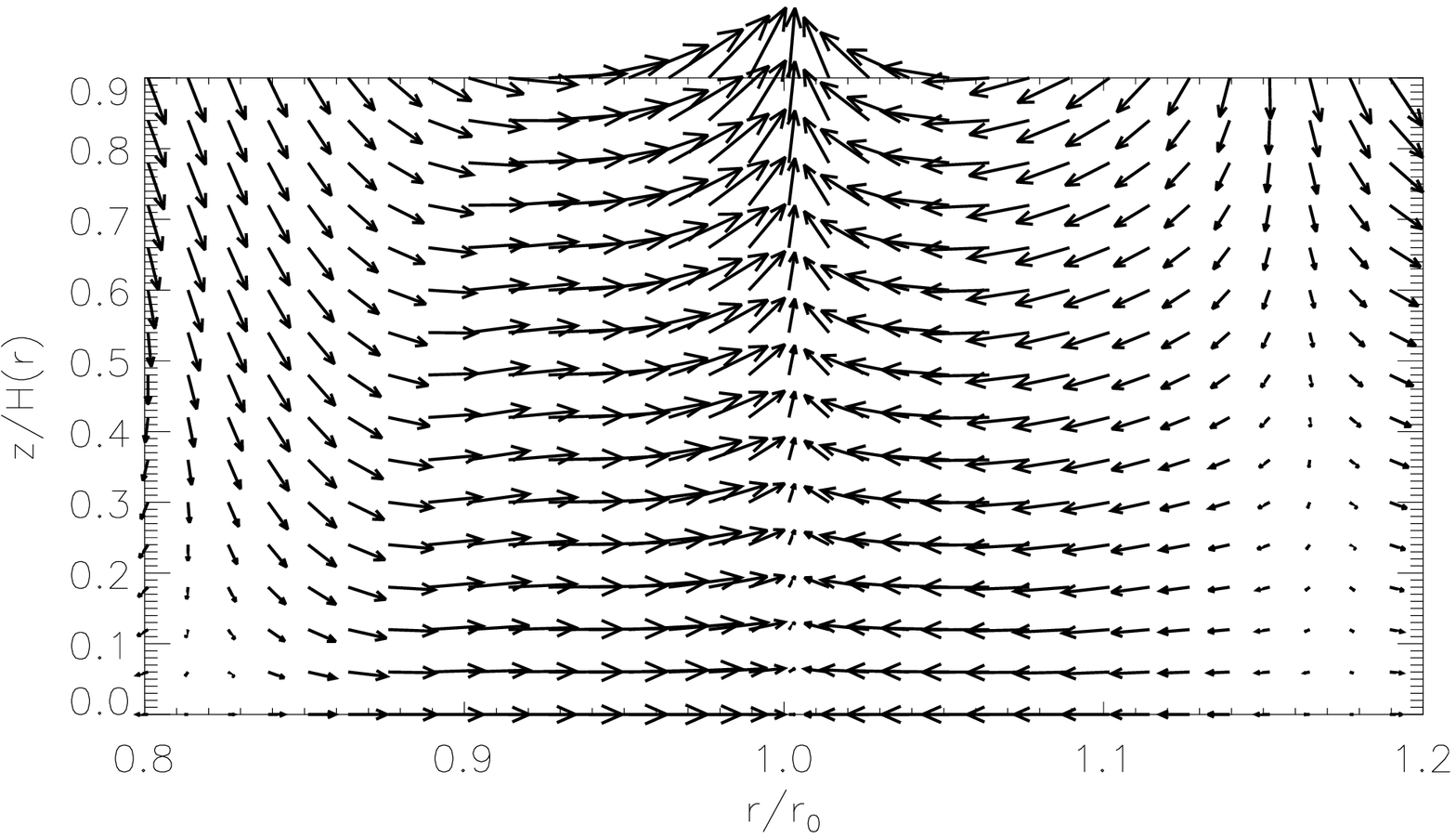} 

   \caption{Reference case with $\Delta P=0$ at $\zmax$. The normalized
     enthalpy perturbation, $|W(r,z)/W(r_0,0)|$  
     (top), and the perturbed meridional velocity 
     (bottom) are shown. The real velocity field is taken at the
     vortex core. Velocities are normalized so its magnitude can be
     compared to other cases.  
     \label{m3_fiducial}}
\end{figure}

The numerical method is first tested against \citetalias{lin12} with 
the open boundary (case 1a).  
Fig. \ref{m3_fiducial} shows $W$ at several heights and the perturbed
real velocity field projected onto the $(r,z)$ plane. 
$|W|$ has a weak vertical dependence in the co-rotation region
$r\in[0.8,1.2]r_0$. The enthalpy perturbation becomes increasingly 
three-dimensional away from $r_0$. Near $\zmax$, the disturbance in
the wave-region can reach amplitudes comparable to that near
$r_0$. This means the RWI becomes radially global at large heights.     

The above result, considered as the intrinsic solution, 
is very similar to \citetalias{lin12}, including the
growth rate and radially converging flow towards $r_0$ with
upwards motion at the vortex core. This validates the choice of
solution method.

\subsection{Solid boundaries}
\begin{figure}
   \centering
   \includegraphics[scale=.425,clip=true,trim=0cm 1cm 0cm
     0cm]{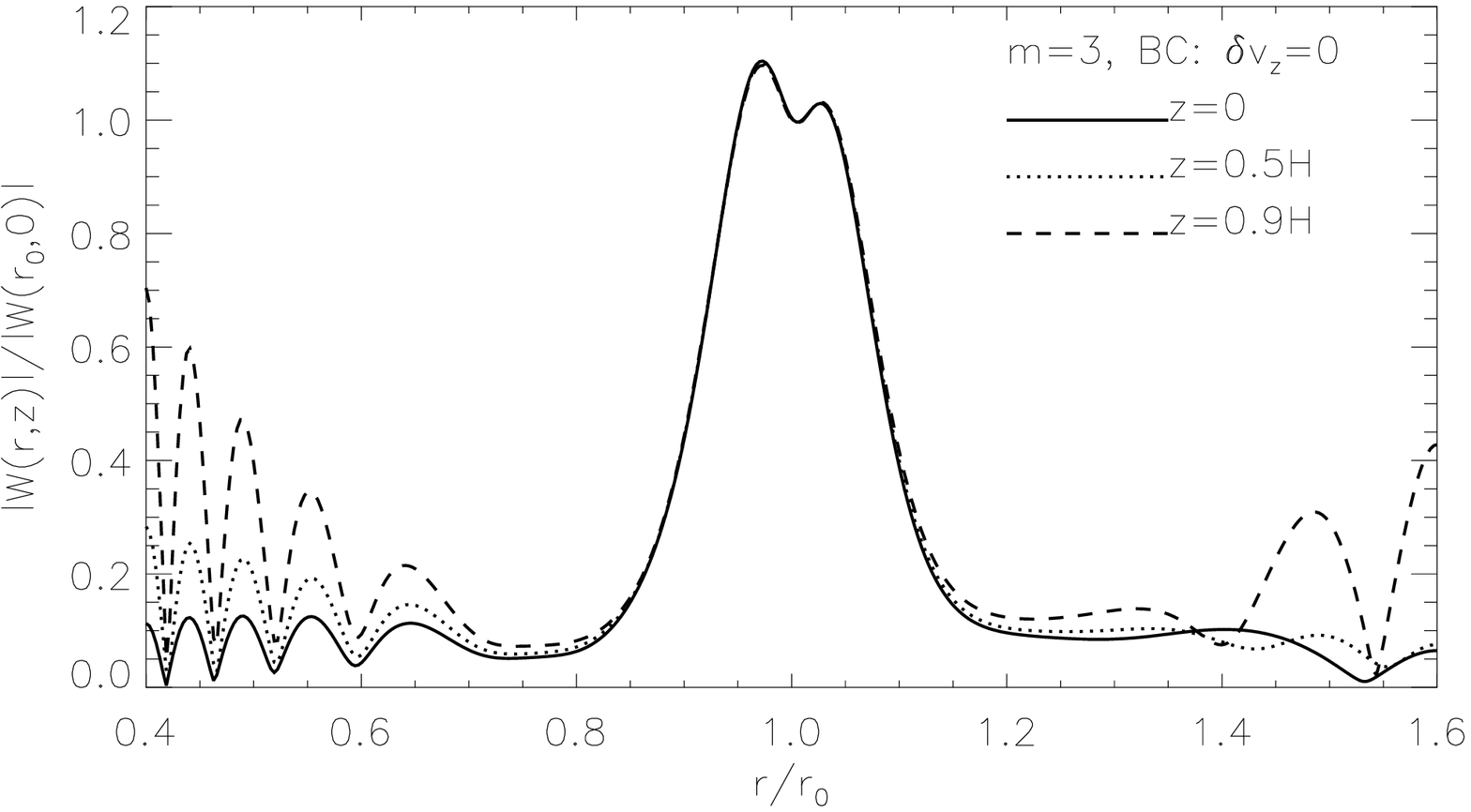}
     \includegraphics[scale=.425,clip=true,trim=0cm 0cm 0cm
       0cm]{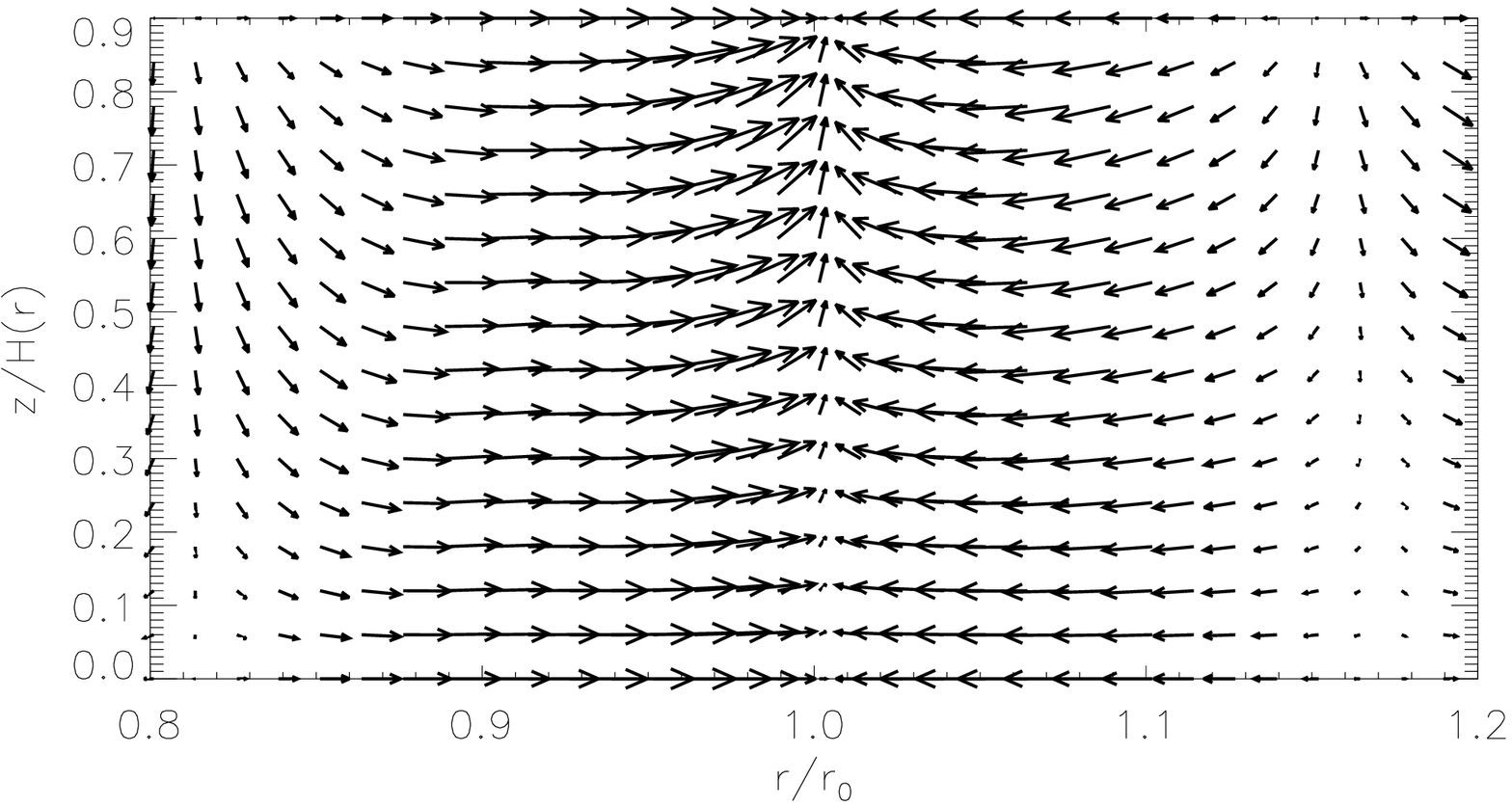} 
   \caption{Same as Fig. \ref{m3_fiducial} but with $\dd v_z=0$
     applied at $\zmax$. 
     \label{m3_vz0}}
\end{figure}

Comparison of case 2a ($\dd v_\perp=0$) and 3a ($\dd v_z=0$)
with the reference case above show that a solid boundary only has a minor
effect on the instability growth rate. The perturbation for
case 3a is plotted in Fig. \ref{m3_vz0}.      

In comparison with case 1a, the enthalpy perturbation in the
co-rotation region is unaffected by solid boundaries. There is, 
however, a decrease in three-dimensionality as indicated by 
$\avg{\theta_m}$. This correlates with a slight increase in 
growth rate, suggesting that suppressing vertical motions favor
instability.       

A solid upper boundary causes the disturbance to become more
two-dimensional in the wave-region. Compared to
Fig. \ref{m3_fiducial}, for $r<0.6r_0$ the amplitude of $W(r,0.9H)$
has decreased while that for $W(r,0)$ has increased. Consequently, the
magnitude of the vertical flow is smaller in case 3a than in case 1a.

\subsection{Effect of domain height}
In case 2b and 3b the vertical domain is reduced from cases 2a
and 3a, respectively. This increases the growth rate by $<8\%$ but 
$\avg{\theta_m}$ is reduced by a factor of 3---4. When $\theta_m$ is
averaged over $Z\in[0,0.45]$, cases 2a and 3a have  
$\avg{\theta_m}=0.19$. So the reduction in three-dimensionality
with decreased $\zmax$ is not simply due to averaging over a region
closer to the midplane (towards which $\dd v_z\to0$). 

\begin{figure}
  \centering
  \includegraphics[scale=.425,clip=true,trim=0cm 0cm 0cm
    0cm]{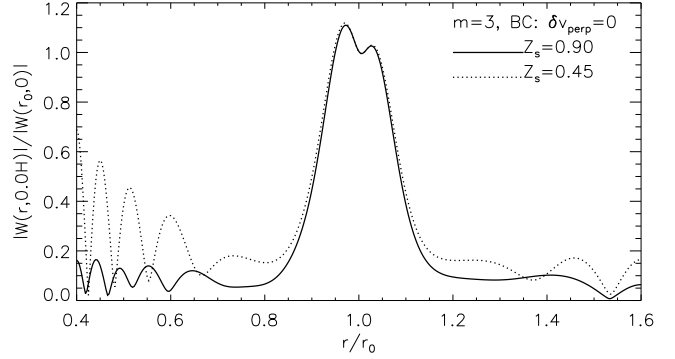} 
  \caption{Midplane enthalpy perturbations subject to $\dd v_\perp=0$ at
    two values of $\zmax$. 
    \label{thickness}}
\end{figure}

Fig. \ref{thickness} compares $|W(r,0)|$ between cases 2a and 2b. The
co-rotation region is unchanged. (Note that $\dd 
v_z\propto \p_ZW/\sbar$ and $|\sbar(r_0)|/\Omega_0\ll 1$, so 
a small difference in $\p_ZW$ between the two cases can still 
result in an appreciable difference in vertical velocity.)   
The perturbation magnitude in $r<0.8,\,r>1.2$ has noticeably increased 
with reduced disc thickness. In this sense the RWI has been made more
radially global by restricting the fluid to a thinner slab.   

Case 1b has  $\Delta P=0$ applied $\zmax=0.45$. This corresponds to
a situation where conditions above the upper boundary adjusts
according to the RWI in $Z<\zmax$, which is of course
unrealistic. Nevertheless, it is interesting to note that 
the growth rate has decreased compared to case 1a, but as
above this correlates with an increase in three-dimensionality.   

\subsection{Dependence on polytropic index}
\begin{figure}
   \centering
   \includegraphics[scale=.425,clip=true,trim=0cm 0cm 0cm
     0cm]{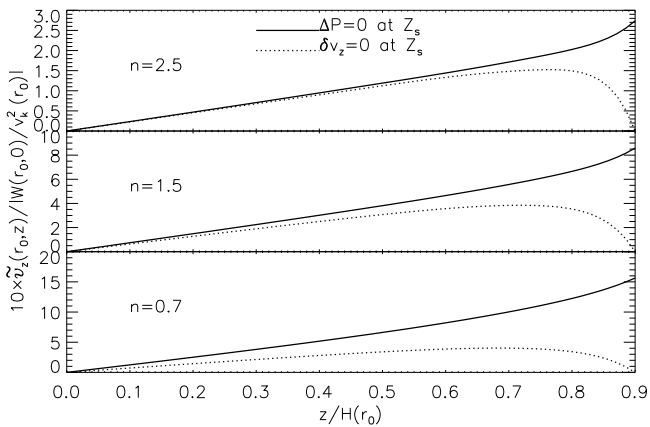} 
   \caption{Normalized vertical velocity near a local maximum in 
     midplane enthalpy perturbation for a range of polytropic indices
     $n$ and upper disc boundary conditions. Here $v_k\equiv r\Omega_k$. 
     \label{varpolyn}}
\end{figure}

\citetalias{lin12} found that the vertical flow in the vortex core
increases in magnitude with decreasing polytropic index when other
parameters are fixed. This is reflected in Fig. \ref{varpolyn} when 
the open boundary is applied. As $n$ is lowered from 2.5 to 0.7, the maximum
vertical velocity increases by a factor of $\sim 6$. 
For each $n$, a calculation with a solid boundary is also shown, for which
the increase in vertical velocity is less significant (by a factor $\sim3$). 

Fig. \ref{varpolyn} indicate upper disc boundary conditions are more
important for smaller polytropic indices.  The dotted
curve for $n=2.5$ only deviates from the solid curve in
$Z\gtrsim0.6$. For $n=0.7$, forcing the flow to be horizontal at
$\zmax$ causes the disturbance to be more two-dimensional throughout
the column of the fluid. 

In previous studies with a vertically isothermal disc (equivalent to a
polytrope with $n\to\infty$), the vortex core appears hydrostatic
\citep[][\citetalias{lin12}]{meheut12}. So for large $n$ explicitly
setting $\dd v_z=0$ at $\zmax$ makes little difference. 
Conversely, as $n$ is lowered the intrinsic solution acquires larger
$|v_z|$ at the vortex core, so that the effect of forcing $\dd
v_z\to0$ at the upper boundary becomes global in the vertical
direction.   

Imagine setting up the RWI with $\Delta P=0$ at $\zmax$, but during
its development force $\dd v_z\to0$ at $\zmax$. Suppose upper boundary effects 
are communicated through the fluid by 
sound waves. The sound-speed in a polytrope is 
$c_s\equiv\sqrt{dP/d\rho}= \Omega_kH(1-Z^2)^{1/2}/\sqrt{2n}$, and the
vertical sound-crossing time is $\tau_c\sim H/c_s\propto
\sqrt{n}/\Omega_k$. In unit time, the change in boundary
condition should affect the lower $n$ polytrope deeper into the disc
than higher $n$, as observed in Fig. \ref{varpolyn}. (This is also seen 
when the bump amplitude $A$ for each $n$ was 
adjusted to achieve equal growth rates.)



\subsection{Other experiments}
Additional calculations have been performed with a variable surface
function $\zmax=\zmax(r)$.  
Examples include a surface of constant aspect-ratio (with $\dd
v_\perp=0$) and a surface of  constant pressure (with $\Delta
P=0$). In this case a second co-ordinate transformation is
required. Similar results to previous
experiments were obtained, including the aforementioned trend of growth rate with
respect to disc thickness. The RWI appears insensitive to the
exact shape of the upper disc boundary as well. 

\section{Discussion}\label{summary}
Linear simulations of the RWI with different upper disc boundary
conditions have been performed. Changing from a free surface to a
solid boundary increase growth rates slightly. This is accompanied by
a decrease in the three-dimensionality of the flow at the vortex core.   
Confining the fluid to a smaller vertical domain has the same
effect. This suggests 3D effects are weakly stabilizing
\citep{li03}. However, upper disc boundary conditions affect the 
vertical flow throughout  the vertical extent of the vortex core. This
effect is stronger for decreasing polytropic index $n$, which
corresponds to decreasing compressibility.          

The above results support previous interpretations that the RWI is largely
two-dimensional \citep{umurhan10,lin12}. The linear growth
rate and general flow pattern is insensitive to details of the upper
disc  boundary conditions, \emph{provided it is consistent with the 2D
solution}. 
A counter-example is vanishing enthalpy 
perturbation at the upper disc boundary\footnote{This 
  can be satisfied by choosing the basis functions as $\psi_k=T_{2k}-T_0$.}. 
Indeed, in this case no RWI-like modes were found (i.e. modes with enthalpy disturbances 
radially confined to the bump). 
Inappropriate upper disc boundary conditions may suppress the RWI as
seen in linear and nonlinear simulations performed to date
\citep[e.g.][]{meheut12b}.    

Experimenting with different vertical boundary conditions, as done here, 
is only a crude way to mimic possible boundary effects in a protoplanetary disc. 
However, failure to find the RWI for some conditions imply that whether or not 
the RWI can develop in protoplanetary discs depends on 
the disc vertical structure, which can be complicated \citep{terquem08}. 
Calculating the RWI for protoplanetary discs will ultimately require 
multi-layer fluid models \citep{umurhan12}. 

The solution method in this paper supersedes \citetalias{lin12} for
its simplicity and ability to admit different upper disc boundary
conditions. Also, it does not rely on the functional form of the 
background stratification. In principle, it can be  
applied to non-barotropic perturbations since inclusion of the energy
equation only changes the coefficients of the governing PDE. 
Assuming the RWI is not modified significantly, then this solution method
should be suitable for discs with entropy gradients. 
This will be the subject of a follow-up paper \citep[][submitted]{lin12b}.

\end{document}